\documentclass[prl,preprint,a4paper,superscriptaddress,citeautoscript,floatfix,showpacs]{revtex4-1}
\usepackage{graphicx}
\usepackage{amsmath}
\usepackage[svgnames]{xcolor}
\usepackage[
	colorlinks=True,linkcolor=DarkRed,citecolor=Blue,urlcolor=MediumBlue,
	pdfstartview=FitH,bookmarks=False,pdfpagemode=UseNone
]{hyperref}
\usepackage{microtype}

\newcommand{\tabref}[1]{Table~\ref{#1}}
\newcommand{\figref}[1]{Fig.~\ref{#1}}

\begin{document}

\title{\boldmath Mass shift of charmonium states in $\bar p A$ collision}

\author{Gy\"orgy Wolf}
\author{G\'abor Balassa}
\affiliation{Institute for Particle and Nuclear Physics, Wigner Research Centre for Physics,
  Hungarian Academy of Sciences, H-1525 Budapest, Hungary}
\email[]{wolf.gyorgy@wigner.mta.hu}
\author{P\'eter Kov\'acs}\
\affiliation{Institute for Particle and Nuclear Physics, Wigner Research Centre for Physics,
  Hungarian Academy of Sciences, H-1525 Budapest, Hungary}
\affiliation{Institute for Theoretical Physics, Johann Wolfgang Goethe University, Max-von-Laue-Str. 1,
60438 Frankfurt am Main, Germany}
\affiliation{ExtreMe Matter Instititute EMMI, GSI Helmholtzzentrum f{\"u}r Schwerionenforschung, Planckstrasse 1, 64291 Darmstadt, Germany}
\author{Mikl\'os Z\'et\'enyi}
\affiliation{Institute for Particle and Nuclear Physics, Wigner Research Centre for Physics,
	Hungarian Academy of Sciences, H-1525 Budapest, Hungary}
\affiliation{ExtreMe Matter Instititute EMMI, GSI Helmholtzzentrum f{\"u}r Schwerionenforschung, Planckstrasse 1, 64291 Darmstadt, Germany}
\author{Su Houng Lee}
\affiliation{Department of Physics and Institute of Physics and Applied Physics, Yonsei University, Seoul 03722, Korea}

\date{\today}

\begin{abstract}
  The masses of the low lying charmonium states, namely, the $J/\Psi$,
  $\Psi(3686)$, and $\Psi(3770)$ are shifted downwards due to the
  second order Stark effect. In $\bar p + \text{Au}$ collisions at
  $6-10$~GeV we study their in\,-\,medium propagation. The time
  evolution of the spectral functions of these charmonium states is
  studied with a Boltzmann\,-\,Uehling\,-\,Uhlenbeck (BUU) type
  transport model. We show that their in\,-\,medium mass shift can be
  observed in the dilepton spectrum. Therefore, by observing the
  dileptonic decay channel of these low lying charmonium states,
  especially for $\Psi(3686)$, we can gain information about the
  magnitude of the gluon condensate in nuclear matter. This
  measurement could be performed at the upcoming PANDA experiment at
  FAIR.
\end{abstract}


\maketitle


\textit{Introduction.---} In Quantum Chromo Dynamics (QCD) the
condensates, like the quark condensate $m_q\langle\bar q q\rangle$
\cite{Shifman:1978bx,Colangelo:2000dp} and the gluon condensate
$\langle\alpha_s G^2\rangle$ \cite{Shifman:1978bx} are fundamental
quantities, which are important to understand hadron
phenomenology. Their values in vacuum are quite well known
\cite{Shifman:1978bx,Shifman:1978by,GellMann:1968rz,Shifman:1978zq,Bertlmann:1983pf,Rakow:2005yn,Narison:2009vy}. However,
in matter we do not have this information, we only know their first
nonzero coefficients in the density expansion
\cite{Drukarev:1988kd,Hatsuda:1991ez,Cohen:1991nk}. The observation of
in\,-\,medium modifications of hadrons may provide us valuable
information about these condensates in matter. While the masses of
hadrons consisting of light quarks changes mainly because of the (partial)
restoration of the chiral symmetry -- through their dependence on the
chiral order parameter $m_q\langle\bar q q\rangle_\rho$ --, those made
of heavy quarks are sensitive mainly to the changes of the non-perturbative
gluon dynamics manifested through the changes in the  gluon
condensates \cite{Luke:1992tm,Klingl:1998sr}. In the low density
approximation the gluon condensate is expected to be reduced by $5-7
\%$ at normal nuclear density
\cite{Meissner:1995,SHLee_Charm}. Therefore, the masses of the
charmonium states -- which can be considered as a color dipole in color
electric field -- are shifted downwards because of the second order
Stark effect \cite{SHLee_Charm,LeeKo,Friman:2002fs}. Moreover, since
the $D$ meson loops contribute to the charmonium self\,-\,energies and
they are slightly modified in\,-\,medium, these modifications generate
further minor contributions to the charmonium in\,-\,medium mass
shifts \cite{LeeKo}.

In this paper, our aim is to propose a way to ``measure'' the gluon
condensate in nuclear matter via studying the mass shifts of the
charmonium states by observing their dileptonic decays.

Antiproton induced reactions are the most prominent candidates to
observe charmed particles in nuclear matter, since the medium is much
simpler in this case than the one created in heavy ion collisions or
even in proton induced reactions. Furthermore, the two main background
contribution to the dilepton yield in the charmonium region, namely
the Drell-Yan and the ``open charm decay'' are expected to be
small. There is only a few energetic hadron\,-\,hadron collisions that
can produce heavy dileptons via the Drell-Yan process.  In the open
charm decay the $D$ mesons decay weakly. The $c$(or $\bar c$) quark in
$D$(or $\bar D$) decays dominantly to leptons and $s$(or $\bar s$)
quark. Consequently, the $e$ and $\bar e$ are usually accompanied by
the $K$ and $\bar K$ mesons. Therefore, not very far above the
threshold, the production of electron\,-\,positron pairs via the open
charm decay with large invariant mass is energetically suppressed. The
observation of charmonium in vacuum and in medium in antiproton
induced reactions is an important goal of the AntiProton Annihilation
at Darmstadt (PANDA) collaboration at the soon to be built Facility
for Antiproton and Ion Research (FAIR) accelerator complex.

The dynamics of the antiproton\,-\,nucleus reactions are described with
a Boltzmann\,-\,Uehling\,-\,Uhlenbeck (BUU) type transport model. The
spectral functions of the $J/\Psi$, $\Psi(3686)$, and $\Psi(3770)$
vector mesons are expected to be modified in a strongly interacting
environment according to \cite{SHLee_Charm,LeeKo,Friman:2002fs}.
Therefore, one has
to propagate the spectral functions of these charmonium states properly.

Similar investigation has been carried out in \cite{Golubeva_Charm},
however, they did not consider substantial in\,-\,medium mass shift
for the charmonium states, {\it i.e.} that work misses the essence of
this investigation. An initial version of this approach is to be
published in \cite{Wolf:2017qbz}, where important ingredients such as
the background contributions and the charmonium absorption were
missing and the collisional broadening was only taken into account
approximately.


\textit{Transport model.---}
Our original model was developed for the energy range of the
  Heavy\,-\,Ion\,-\,Synchrotron (SIS18) experiment at GSI in Germany.
It contained 27 baryons and 6 mesons. The details of this transport
model can be found in \cite{KampferWolf-2010,Wolf90,Wolf93}.

Recently, we improved the model in order to be applicable for higher
energies. The relevant changes concern on the built\,-\,in elementary
cross sections of the model, namely, we included cross sections for
the production of charmonium and D\,-\,mesons states. We calculated
these unknown cross sections, such as $\bar p p \to J/\psi \pi$, or
$\bar p p \to D\bar D$ with the help of a statistical bootstrap model
developed by some of us \cite{Balassa:2017pgf}. The
antiproton\,-\,nucleon cross section was set to $20$~mb taken from
\cite{Landolt-Bornstein}. We apply energy independent charmonium
absorption cross section for every hadrons as $4.18$~mb for $J/\Psi$
and $7.6$~mb for $\Psi(3686)$ and $\Psi(3770)$ according to Ref.
\cite{Linnyk:2006ti}. In $\bar p A$ collisions at relativistic energies
charmonium absorption does not play such an important role as at
  ultrarelativistic energies, since the hadron density is much less
here. It should be noted that the charmonium states in the transport
model are produced perturbatively. That is, after they are created with
some probability through the antiproton annihilation process, in every
time step it could decay with some probability, however, we do not let
it decay, instead, we use this probability to add its contribution to
the dilepton spectra.

If we create a particle in a medium with an in-medium mass, through
its evolution, it should regain its vacuum mass, when it leaves the
collision zone. If a local density approximation is used for changing
its mass, the energy conservation will be clearly violated. For the
propagation of off\,-\,shell particles a more sophisticated method is
needed. One can describe the in\,-\,medium properties of particles
with a so\,-\,called ``off-shell transport''. These equations are
derived by starting from the Kadanoff\,-\,Baym equations
\cite{Baym:1961zz} for the Green's functions of the
particles. Applying first\,-\,order gradient expansion after a Wigner
transformation \cite{Cassing-Juchem00,Leupold00} one arrives at a
transport equation for the retarded Green's function. To solve
numerically the Kadanoff\,-\,Baym equations one may exploit the
test\,-\,particle ansatz for the retarded Green's function
\cite{Cassing-Juchem00,Leupold00}.

The equations of motion of the test\,-\,particles have to be supplemented
by a collision term, which creates couplings among the equations for the various
particle species. It can be shown \cite{Leupold00} that the collision
term has the same form as in the standard BUU approach. The same
model was used to study the propagation of low mass vector meson
spectral functions at lower energies \cite{KampferWolf-2010,Almasi}.

The explicit form of the ``off-shell transport'' equations can
  be found in \cite{Cassing-Juchem00,Wolf:2017qbz}. To solve those
  equations an explicit expression for the real and imaginary part of
  the self\,-\,energy of the charmonium particle at hand is needed. In
  our calculations the following simple, density dependent form is
  assumed for each charmonium state -- indexed by $V$,
\begin{eqnarray} 
{\rm \Re} \Sigma^{ret}_V &=& 2 m_V \Delta m_V \frac{\rho}{\rho_0}, \label{areal}\\
{\rm \Im} \Sigma^{ret}_V &=& -m_V (\Gamma^{\text{vac}}_V + \Gamma_{\text{coll}}). \label{aimag}
\end{eqnarray}
Eq.~\eqref{areal} results in a ``mass shift'' of the form
  $\Delta m^{\text{shift}}_V =
  \sqrt{m_V^2+\Re\Sigma_V^{ret}}-m_V\approx \Delta m_V
  \frac{\rho}{\rho_0}$, where $\rho_0 = 0.16 \frac{1}{\text{fm}^3}$ stands for
  the normal nuclear density. The imaginary part incorporates a vacuum
width $\Gamma^{\text{vac}}_V$ term and a collisional broadening term having the form
\begin{equation}
  \Gamma_{\text{coll}} = \frac{v \sigma \rho}{\sqrt{(1-v^2)}},
\end{equation}
where $v$ is the relative velocity of the particle in the local rest
frame, $\sigma$ is the total cross section of the particle colliding
with nucleons and $\rho$ is the local density. The parameters ($\Delta
m_V)$ are taken from \cite{LeeKo} and are given in
\tabref{Tab:param_dmV}. The first values in \tabref{Tab:param_dmV}
come from the second order Stark\,-\,effect (which depends on the
gluon condensate), while the second ones emanate from the D\,-\,meson
loops.

\begin{table}[th]
  \caption{\label{Tab:param_dmV} Charmonium mass shift parameter
    values taken from \cite{LeeKo}. In $\Delta m_V$ the first terms
    result from the second order Stark\,-\,effect, while the second
    ones emanate from the D\,-\,meson loops.}
  \begin{ruledtabular}
  \begin{tabular}{cc}
    Charmonium type ($V$) &  $\Delta m_V$\\
    \hline
    $J/\Psi$     & $-8+3$~MeV  \\
    $\Psi(3686)$ & $-100-30$~MeV \\
    $\Psi(3770)$ & $-140+15$~MeV \\
  \end{tabular}
  \end{ruledtabular}  
\end{table}

The electromagnetic branching ratios $Br^{\text{el}}$ of the
charmonium states decaying into dileptons may change due to the
broadening. In our approximation the electromagnetic in\,-\,medium
widths are kept at the same value as in vacuum
($\Gamma_{\rho}^{\text{el}} = \Gamma_{\text{vac}}^{\text{el}}$), {\it
  i.e.} they are not increased by the collisional
broadening. Consequently, the electromagnetic in\,-\,medium branching
ratio will change in line with the changes of the total decay width,
$Br_{\rho}^{\text{el}} = \Gamma_{\text{vac}}^{\text{el}}/\Gamma_{\rho}^{\text{tot}}$. The
in\,-\,medium electromagnetic widths are probably larger than the
vacuum ones, but to be on the safe side the minimal value is chosen
not to overestimate the resulting dilepton invariant mass spectrum in
comparison with the background, that is, the genuine change in
$\Gamma_{\rho}^{\text{el}}$ could only increase the spectrum not
decrease.

If a meson is generated at a given density, its mass is distributed in
accordance with its in\,-\,medium spectral function. If the given
meson propagates into a region of higher (or lower) density, then its
mass will decrease (or increase). This method, which is based on
the "off\,-\,shell" transport equations, is energy conserving. We note
that the propagation of $\omega$ and $\rho$ mesons
in the energy range of the HADES experiment at GSI ($\sim
  1-2$~AGev) have been investigated in \cite{KampferWolf-2010} with
the same method.


\textit{Results.---}
\begin{figure}[th]
  \centering
  \includegraphics[width=0.9\columnwidth]{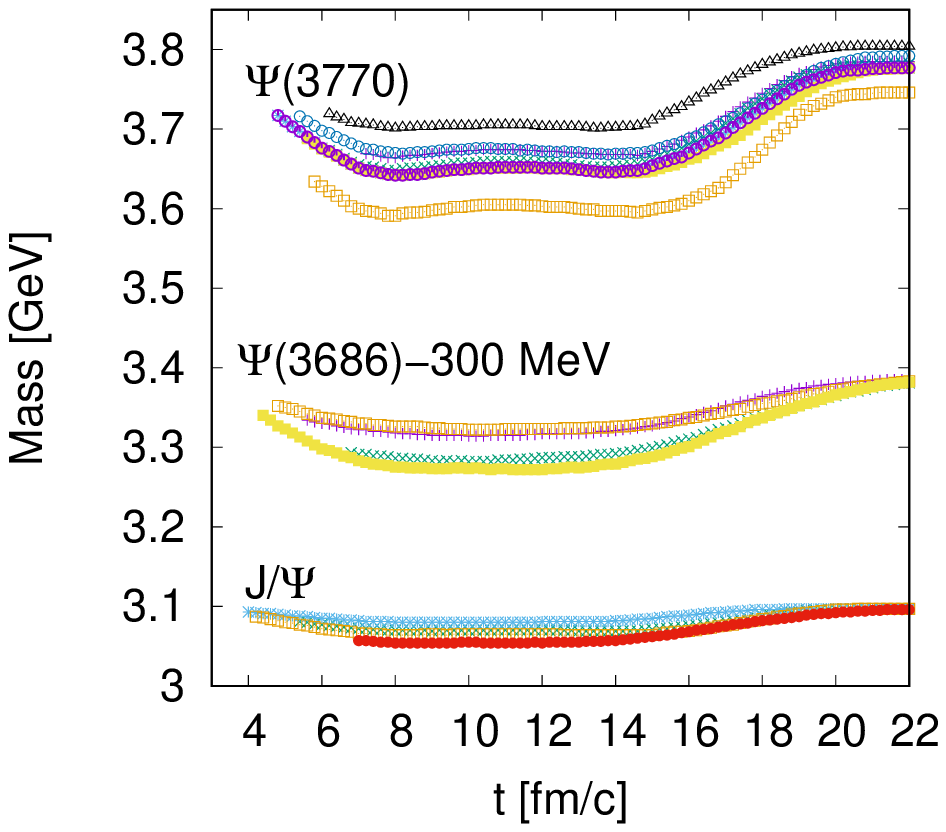}
  \includegraphics[width=0.9\columnwidth]{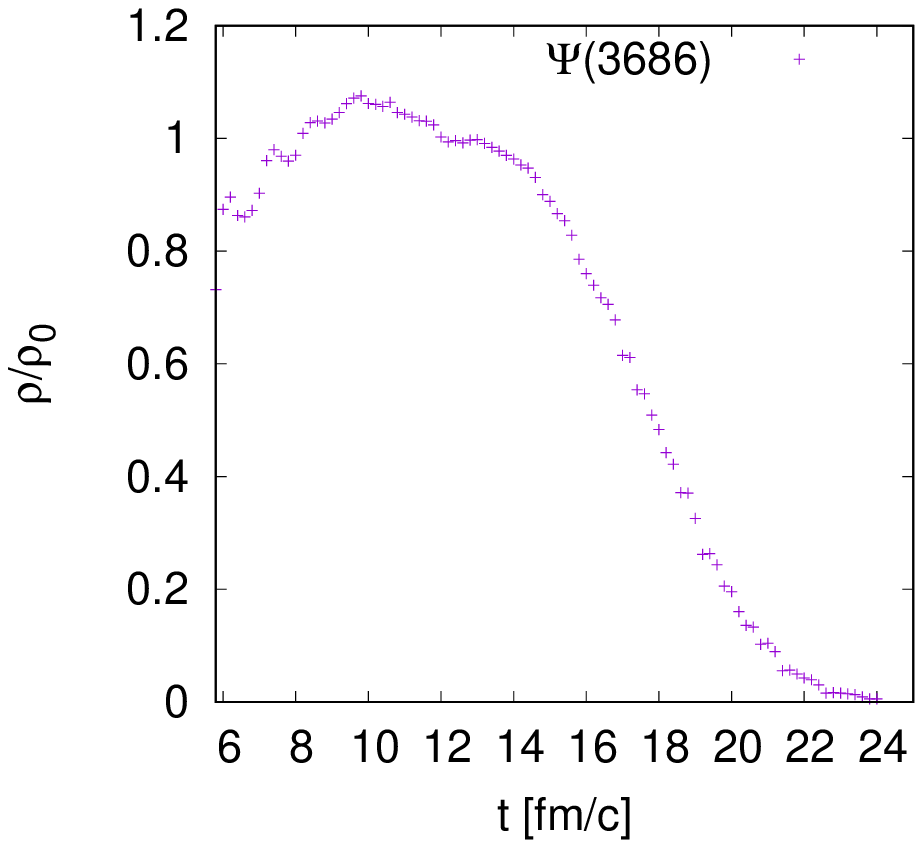}
    \caption{Top: Evolution of the masses of the $J/\Psi$,
      $\Psi(3686)$, and $\Psi(3770)$ charmonium states. For the better
      visibility the masses of the $\Psi(3686)$ are shifted downwards
      artificially by 300 MeV. Each line belonging to a given
        charmonium state represents a different individual particle
        evolution through the medium. Bottom: The average density
      felt by the charmonium states as a function of time.}
    \label{Fig:Charmonium_evol}
\end{figure}
In the top panel of \figref{Fig:Charmonium_evol} the evolution of the
masses of randomly chosen test particles representing charmonium mesons is
shown in case of $\bar p + \text{Au}$ collisions at $6$~GeV bombarding energy.  It
should be noted that for getting a better overview in the figure the
mass of $\Psi(3686)$ was shifted downwards by $300$~MeV. It can be
seen that at the end of the collision process, where the density is
very low, the masses reach the vacuum value as expected. However,
the mass of the $\Psi(3770)$ state spreads even at the
end of the collision due to its substantial vacuum width. Most of the
time the masses of these mesons are either shifted downwards to the
mass corresponding to a density around $\rho_0$ or at their vacuum
value, {\it i.e.} there are only relatively short transition
periods in between. This is because the surface layer is quite narrow
compared to the diameter of the gold nucleus. The evolution of the average
density felt by the charmonium states are shown in the bottom panel of
\figref{Fig:Charmonium_evol}. The same conclusion can be drawn based
on this figure too, namely, the transition period from the dense
medium to the vacuum is small, approximately $4$~fm/c long.

\begin{figure}[th]
  \centering
  \includegraphics[width=0.9\columnwidth]{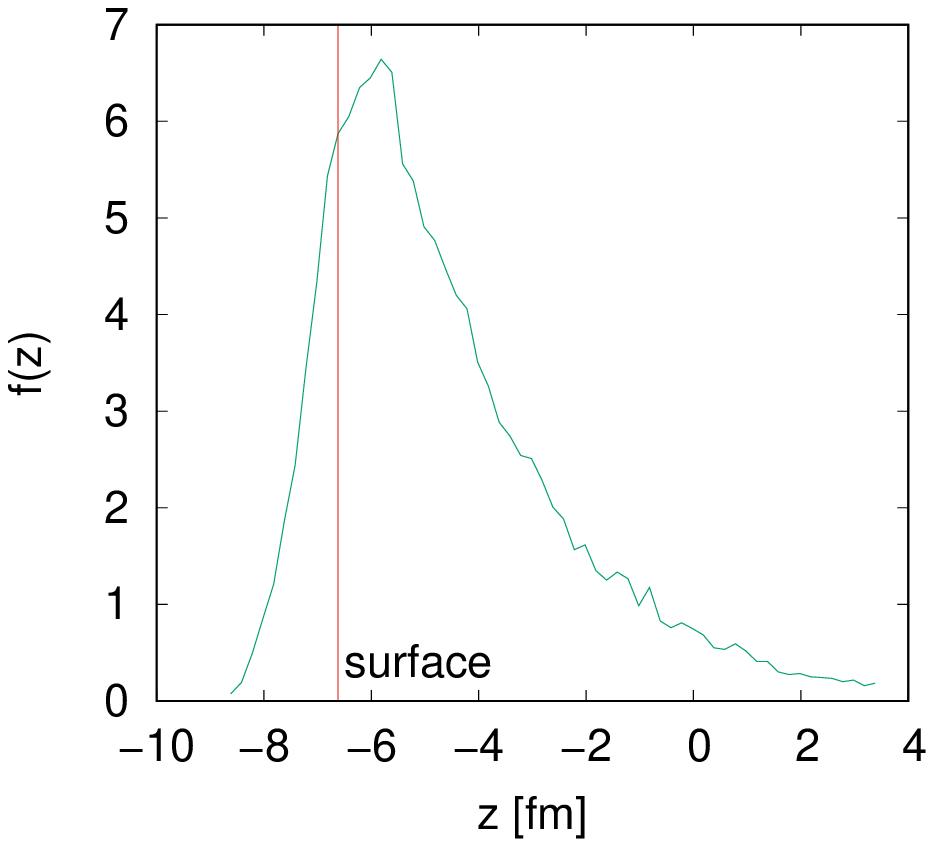}
  \includegraphics[width=0.9\columnwidth]{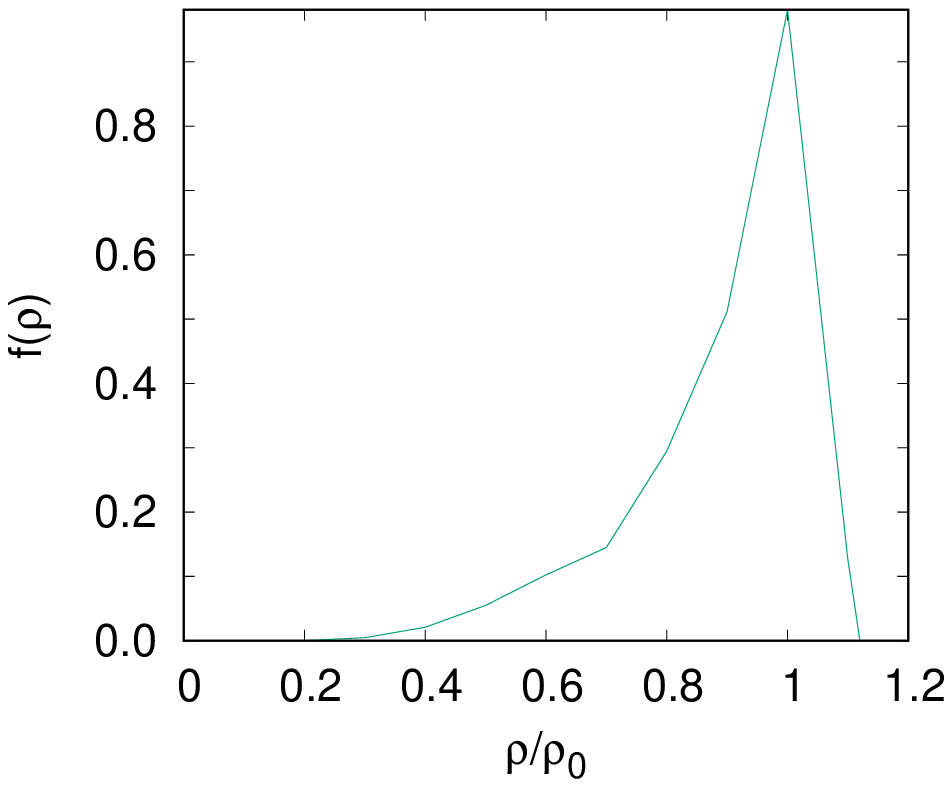}
  \caption{Top: Distribution of the z coordinate of the annihilation
    position of the antiproton, where the z axis is in parallel with
    the beam momentum. The vertical line shows the surface of the nucleus
    on the side of the projectile. Bottom: The density distribution of the
    charmonium creation points.}
    \label{Fig:Charmonium_create}
\end{figure}
In \figref{Fig:Charmonium_create} the antiproton penetration
  into the nucleus is shown. In the top figure we show that there
are some annihilation even before the $\bar p$ reaches the surface of
the nucleus due to the size of the projectile and to the surface thickness
of the target ($\approx \pm 0.5$~fm). Most of the antiprotons annihilate
close to the surface but a substantial amount can reach the center of
the target, since at these energies the annihilation cross section is
not too large ($\approx 20$~mb). In the bottom figure we show the
density distribution of the annihilation points. Most of the
charmoniums are created at densities close to $\rho_0$.

To summarize the dynamics of charmonium production the
  following plausible picture can be given: Most of the antiprotons
annihilate on, or close to the surface of the heavy nucleus creating
a charmonium (with some probability). The charmonium travels
through the interior of the nucleus giving some contribution to the
dilepton yield. That is the charmoniums are treated
  perturbatively (see also the 'Transport model' section).
Traversing the thin surface again on the other side of the
  nucleus, it arrives to the vacuum, where most of the charmonium
actually decays.
\begin{figure}[tb]
  \centering
  \includegraphics[width=\columnwidth]{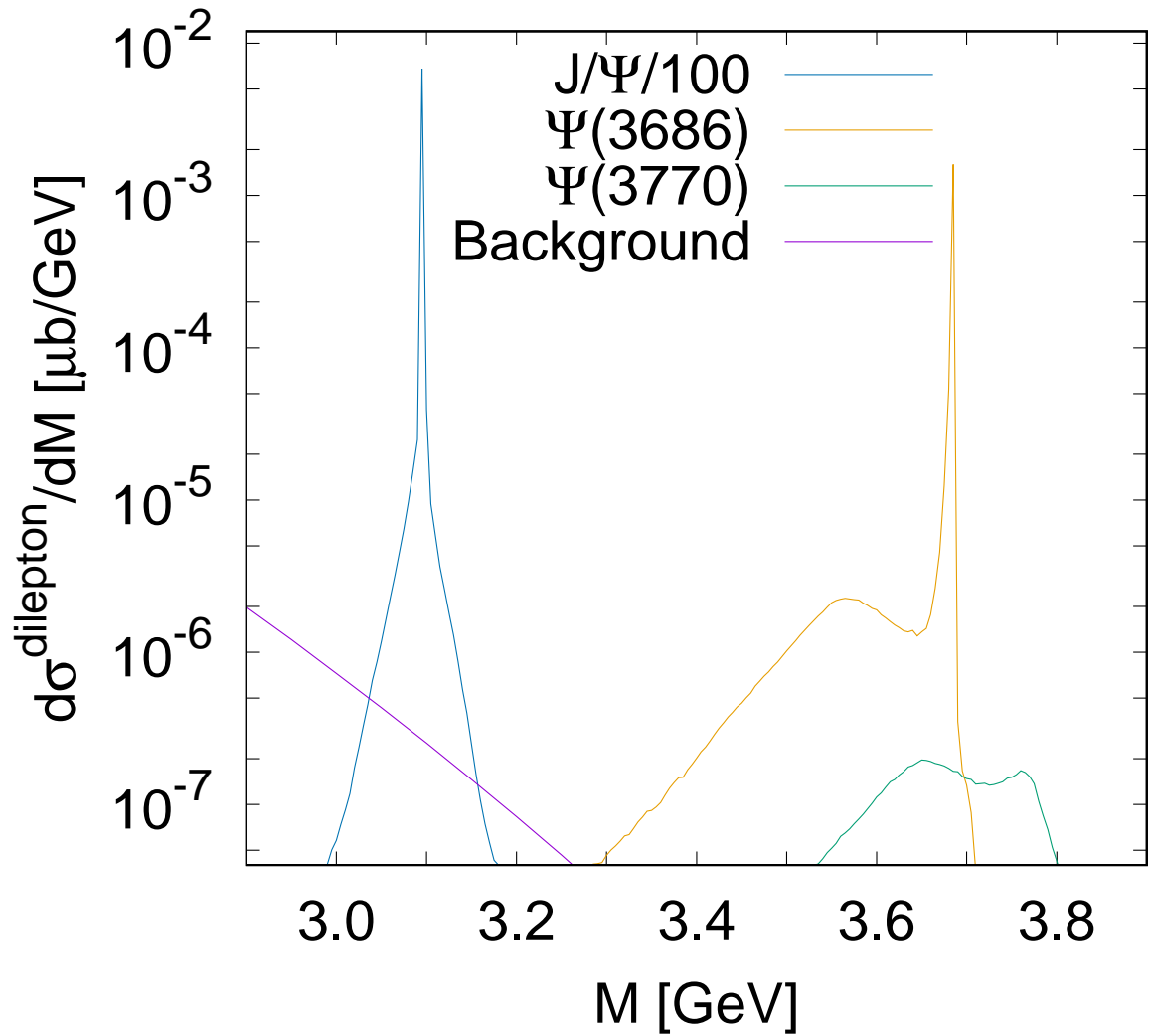}
  \caption{Charmonium contribution to the dilepton spectra in a $\bar p \text{Au}$
    collision at 6 GeV bombarding energies, where the
    in\,-\,medium modifications are accounted for. The background is
    the sum of the Drell\,-\,Yan and the open charm contributions,
    while the impact parameter is integrated from zero to $4.5$~fm.
  The contribution of $J/\Psi$ is divided by 100 to get a better overview.}
  \label{Fig:Dilepspectra}
\end{figure}

In Fig.~\ref{Fig:Dilepspectra} the charmonium contributions to the
dilepton spectrum in a central $\bar p + \text{Au}$ is shown in an
$6$~GeV collision. The impact parameter is integrated out from zero to
$4.5$~fm, which gives approximately the central one third of the total cross
section (radius of the Au nucleus is about $6.6$~fm).

The dilepton invariant mass spectra for the three mentioned
charmonium states are shown along with the background contribution,
which is the sum of Drell\,-\,Yan and open charm decays. Since in our
calculations detector resolution is not included, the vacuum
contributions for the $J/\Psi$ and for the $\Psi(3686)$ result in very
sharp peaks, almost discrete lines. A two\,-\,peak structure can be
seen for each meson, corresponding to its vacuum and the in-medium mass from
the interrior of the nucleus.
However, for the $J/\Psi$ the effect is
negligible, since its mass shift is very small. For the $\Psi(3770)$
there are two peaks, however, its yield is covered by the contribution
of $\Psi(3686)$, thus its observation could be very difficult. On the
other hand, in the $\Psi(3686)$ dilepton mass spectra, the two peaks
are clearly observable. The peak from the medium contribution is
around two times higher than the deepest point in the valley between
the peaks (note the logarithmic scale).
Consequently, the two peaks should be
experimentally easily separated and observed. The shift between the
peaks corresponds approximately to a mass shift at $0.9 \rho_0$
density. We repeated the calculation with double and half mass shifts,
too. The qualitative picture has not changed. Considering $\Psi(3686)$
the two peak structure has remained, and the distance between the
peaks has also been corresponded to the mass shift at the same
$0.9 \rho_0$ density. This gives us a chance to determine the
value of the gluon condensate in nuclear matter, at around
$0.9\rho_0$ density, by measuring the distance between the two
peaks for $\Psi(3686)$, since we know the dependence
of the mass shift of the charmonium on the gluon condensate.
We performed the same calculation
for antiproton bombarding energies $8$ and $10$~GeV as well, which
gave qualitatively the same result. At even lower energies, the peak
structures are distorted strongly, thus the effect is not cleanly
visible. Probably, $6$ GeV is the best bombarding energy, since at higher
energies the annihilation cross section is lower, so more antiprotons penetrate
deep inside the target, giving less contribution from the dense region to
the dilepton yield. At higher energies the background is also higher.

The double peak structure of charmonium contributions to the dilepton
invariant mass spectrum is a novel feature of our model in contrast to
the work of Ref.\ \cite{Golubeva_Charm} where the mass shift of the charmonium
states was not considered..


\textit{Summary.---} We calculated the charmonium contribution to the
dilepton invariant mass spectra. We have shown that via their
dileptonic decay there is a good chance to observe the in-medium
modification of the higher charmonium state $\Psi(3686)$ in a central
$\bar p + \text{Au}$ $6$~GeV collision. This opens up the unique
possibility to ``measure'' the gluon condensate in nuclear matter.
The distance of the two peaks corresponds to a mass shift at
approximately $0.9 \rho_0$ density. The D\,-\,meson loop contributes only by
$25-30$~MeV to the mass shift. The rest (which is expected to
be the major part) is the result of the second order Stark effect,
thus we can determine the gluon condensate that has resulted in
  such a mass shift. Therefore $\bar p A$ collision could provide us
valuable information on the in\,-\,medium properties of the strong
interaction.

The considered energy regime will be available by the
  forthcoming PANDA experiment at FAIR.


\textit{Acknowledgments.---} Gy.~W., M.~Z., G.~B., and P.~K. were
supported by the Hungarian OTKA fund K109462 and Gy.~W., M.~Z. and
P.~K. by the HIC for FAIR Guest Funds of the Goethe University
Frankfurt. P.~K. and M.~Z. also acknowledge support from the 
ExtreMe Matter Institute EMMI at the GSI Helmholtzzentrum f\"ur 
Schwerionenforschung, Darmstadt, Germany. The work of SHL was supported by
the Korea National Research Foundation under the grant number
2016R1D1A1B03930089.







\begin{thebibliography}{99}

\bibitem{Shifman:1978bx} 
  M.~A.~Shifman, A.~I.~Vainshtein and V.~I.~Zakharov,
  Nucl.\ Phys.\ B {\bf 147}, 385 (1979).

\bibitem{Colangelo:2000dp} 
  P.~Colangelo and A.~Khodjamirian,
  In *Shifman, M. (ed.): At the frontier of particle physics, vol. 3* 1495-1576.

\bibitem{Shifman:1978by} 
  M.~A.~Shifman, A.~I.~Vainshtein and V.~I.~Zakharov,
  Nucl.\ Phys.\ B {\bf 147}, 448 (1979).

\bibitem{GellMann:1968rz} 
  M.~Gell-Mann, R.~J.~Oakes and B.~Renner,
  Phys.\ Rev.\  {\bf 175}, 2195 (1968).
  
\bibitem{Shifman:1978zq} 
  M.~A.~Shifman, A.~I.~Vainshtein, M.~B.~Voloshin and V.~I.~Zakharov,
  Phys.\ Lett.\  {\bf 77B}, 80 (1978).

\bibitem{Bertlmann:1983pf} 
  R.~A.~Bertlmann and J.~S.~Bell,
  Nucl.\ Phys.\ B {\bf 227}, 435 (1983).

\bibitem{Rakow:2005yn} 
  P.~E.~L.~Rakow,
  PoS LAT {\bf 2005}, 284 (2006).
  
\bibitem{Narison:2009vy} 
  S.~Narison,
  Phys.\ Lett.\ B {\bf 673}, 30 (2009).

\bibitem{Drukarev:1988kd} 
  E.~G.~Drukarev and E.~M.~Levin,
  Nucl.\ Phys.\ A {\bf 511}, 679 (1990)
  Erratum: [Nucl.\ Phys.\ A {\bf 516}, 715 (1990)].

\bibitem{Hatsuda:1991ez}  
  T.~Hatsuda and S.~H.~Lee,
  Phys.\ Rev.\ C {\bf 46}, no. 1, R34 (1992).

\bibitem{Cohen:1991nk} 
  T.~D.~Cohen, R.~J.~Furnstahl and D.~K.~Griegel,
  Phys.\ Rev.\ C {\bf 45}, 1881 (1992).

\bibitem{Luke:1992tm}  
  M.~E.~Luke, A.~V.~Manohar and M.~J.~Savage,
  Phys.\ Lett.\ B {\bf 288}, 355 (1992).

\bibitem{Klingl:1998sr} 
  F.~Klingl, S.~s.~Kim, S.~H.~Lee, P.~Morath and W.~Weise,
  Phys.\ Rev.\ Lett.\  {\bf 82}, 3396 (1999)
  Erratum: [Phys.\ Rev.\ Lett.\  {\bf 83}, 4224 (1999)].
  
\bibitem{Meissner:1995} 
  B.~Borasoy, U.~G.~Meißner,
  Phys.\ Lett.\ B {\bf 365}, 285 (1995)
  
\bibitem{SHLee_Charm} 
  S.~H.~Lee,
  AIP Conf.\ Proc.\  {\bf 717}, 780 (2004);
  K.~Morita and S.~H.~Lee,
  Phys.\ Rev.\ C {\bf 85}, 044917 (2012).
  
\bibitem{LeeKo} 
  S.~H.~Lee and C.~M.~Ko,
  Phys.\ Rev.\ C {\bf 67}, 038202 (2003).

\bibitem{Friman:2002fs} 
  B.~Friman, S.~H.~Lee and T.~Song,
  Phys.\ Lett.\ B {\bf 548}, 153 (2002)
    
\bibitem{Golubeva_Charm}
  Y.~S.~Golubeva, E.~L.~Bratkovskaya, W.~Cassing and L.~A.~Kondratyuk,
  Eur.\ Phys.\ J.\ A {\bf 17}, 275 (2003)

\bibitem{Wolf:2017qbz} 
  G.~Wolf, G.~Balassa, P.~Kov\'acs, M.~Z\'et\'enyi and S.~H.~Lee,
  Acta Phys.\ Polon.\ Supp. (to be published).

\bibitem{KampferWolf-2010}
  H.~W.~Barz, B.~Kampfer, G.~Wolf and M.~Zetenyi,
  Open Nucl.\ Part.\ Phys.\ J.\  {\bf 3}, 1 (2010);
  G.~Wolf, B.~Kampfer and M.~Zetenyi,
  Phys.\ Atom.\ Nucl.\  {\bf 75} (2012) 718.

\bibitem{Wolf90}
  G.~Wolf, G.~Batko, W.~Cassing, U.~Mosel, K.~Niita and M.~Schaefer,
  Nucl.\ Phys.\ A {\bf 517}, 615 (1990).

\bibitem{Wolf93}
  G.~Wolf, W.~Cassing and U.~Mosel,
  Nucl.\ Phys.\ A {\bf 552}, 549 (1993);
  S.~Teis, W.~Cassing, M.~Effenberger, A.~Hombach, U.~Mosel and G.~Wolf,
  Z.\ Phys.\ A {\bf 356}, 421 (1997).
  
\bibitem{Balassa:2017pgf} 
  G.~Balassa, P.~Kov\'acs and G.~Wolf,
  Eur.\ Phys.\ J.\ A (to be published).
  
\bibitem{Landolt-Bornstein} 
  A.~Baldini, V.~Flaminio, W.~G.~Moorhead, D.~R.~O.~Morrison and
  H.~Schopper,
  BERLIN, GERMANY: SPRINGER (1988) 409 P. (LANDOLT-BOERNSTEIN. NEW SERIES, 1/12B)
  
\bibitem{Linnyk:2006ti} 
  O.~Linnyk, E.~L.~Bratkovskaya, W.~Cassing and H.~Stoecker,
  Nucl.\ Phys.\ A {\bf 786}, 183 (2007).

\bibitem{Baym:1961zz} 
  G.~Baym and L.~P.~Kadanoff,
  Phys.\ Rev.\  {\bf 124}, 287 (1961).
  
\bibitem{Cassing-Juchem00}
  W.~Cassing and S.~Juchem,
  Nucl.\ Phys.\ A {\bf 672}, 417 (2000).

\bibitem{Leupold00}
  S.~Leupold,
  Nucl.\ Phys.\ A {\bf 672}, 475 (2000).

\bibitem{Almasi} 
  G.~A.~Alm\'asi and G.~Wolf,
  Nucl.\ Phys.\ A {\bf 943}, 117 (2015).

\end{thebibliography}
\end{document}